\documentclass[twocolumn,showpacs,preprintnumbers,amsmath,amssymb,prl]{revtex4}

\usepackage[dvips]{graphicx}

\usepackage{dcolumn}
\usepackage{bm}

\begin{document}

\title{
Time domain Einstein-Podolsky-Rosen correlation }

\author{Nobuyuki Takei$^{1,2}$}%
 \email{takei@alice.t.u-tokyo.ac.jp}
\author{Noriyuki Lee$^{1,2}$}
\author{Daiki Moriyama$^{1}$}
\author{J. S. Neergaard-Nielsen$^{1,3}$}
\author{Akira Furusawa$^{1,2}$}

\affiliation{
$^{1}$Department of Applied Physics, School of Engineering, The University of Tokyo, \\
7-3-1 Hongo, Bunkyo-ku, Tokyo 113-8656, Japan\\
$^{2}$CREST, Japan Science and Technology Agency, 1-9-9 Yaesu, Chuo-ku, Tokyo 103-0028, Japan\\
$^{3}$Niels Bohr Institute, University of Copenhagen, DK 2100, Denmark
}%

\date{\today}

\begin{abstract}

We experimentally demonstrate creation and characterization of
Einstein-Podolsky-Rosen (EPR) correlation between optical beams in the time
domain. The correlated beams are created with two independent
continuous-wave optical parametric oscillators and a half beam splitter. We
define temporal modes using a square temporal filter with duration $T$ and
make time-resolved measurement on the generated state. We observe the
correlations between the relevant conjugate variables in time domain which
correspond to the EPR correlation. Our scheme is extendable to continuous
variable quantum teleportation of a non-Gaussian state defined in the time
domain such as a Schr\"odinger cat-like state.

\end{abstract}

\pacs{03.67.Mn, 03.65.Ud, 42.50.Dv}

\maketitle

Quantum entanglement is one of the most fundamental properties of quantum
mechanics and also the essential ingredient in quantum information
processing~\cite{NC00,Braunstein05}. The nonclassical correlation was first
argued by Einstein, Podolsky, and Rosen~(EPR) relating to canonically
conjugate continuous variables~(CVs)~\cite{EPR35}. This originally devised
EPR correlation has been experimentally realized by employing a two-mode
squeezed vacuum state~\cite{Reid89,Ou92}, where the relevant variables are
quadrature-phase amplitudes of optical fields. Since this significant
milestone, there have been further experimental observations of the EPR
correlations~\cite{Silberhorn01,Schori02,Bowen03,Wenger05} and theoretical
characterizations~\cite{Duan00,Simon00}. Moreover the EPR correlation, or
the two-mode squeezed vacuum has been used as a resource in implementation
of quantum protocols. Thanks to well-established tools of manipulating the
EPR correlation, CV quantum information processing~(QIP) has been rapidly
developing in recent years~\cite{Braunstein05} and has been successful in
implementing quantum protocols such as quantum
teleportation~\cite{Braunstein98,Furusawa98,Takei05}.

Quantum teleportation is one of the fundamental quantum protocols which
exploit the EPR correlation or quantum
entanglement~\cite{Bennett93,Braunstein98}. By using an optical field, it
has been experimentally realized for Gaussian input states; a coherent
state~\cite{Furusawa98,Takei05}, a squeezed state~\cite{Takei05b}, and an
EPR state, i.e., entanglement swapping~\cite{Takei05}. These successes are
based on well-developed techniques of optical Gaussian operations
consisting of beam splitters, phase shifting, squeezing, phase-space
displacement and homodyne detection. In addition to the teleportation, CV
quantum protocols have so far been performed only with Gaussian states and
operations. However non-Gaussian states or non-Gaussian operations are
further required to extract versatile potential of CV QIP~\cite{Lloyd99}.
Thus quantum teleportation of a non-Gaussian state would become the next
important challenge toward universal CV QIP.

For the generation of a non-Gaussian state in an optical system, there is
considerable interest in the techniques  where photon counting is performed
for a tapped-off beam from a Gaussian state light beam. When counting a
photon, the tapped Gaussian state is converted to a non-Gaussian state, a
technique called  ``photon-subtraction". Recently there are some reports
which show the generation of non-Gaussian states (Schr\"odinger cat-like
states) with this technique~\cite{Ourjoumtsev06,Jonas06}. Since the photon
counting event is defined in the time domain, conventional CV teleportation
schemes with a sideband of an optical carrier \cite{Furusawa98} cannot be
used for the teleportation of such states. Therefore, in order to realize
CV teleportation of such non-Gaussian states, one need to generate EPR
correlation in the time domain.

In this Letter we demonstrate experimental generation and characterization
of a two-mode squeezed vacuum state in the time domain, a step towards CV quantum
teleportation of non-Gaussian states like a Schr\"odinger cat-like state.
The main feature of our EPR resource is that we use two squeezed vacuum
states of continuous-wave (CW) light beams from two independent
subthreshold optical parametric oscillators~(OPOs) to obtain well-defined
frequency and spatial modes, and also that we utilize almost the whole
frequency bandwidth of the OPO cavities to define a quantum state in the
time domain. The use of the broad bandwidth allows us to make time-resolved
measurements like photon counting as described below. The experiment
presented here is quite different from the previous works in the
frequency domain~\cite{Ou92,Silberhorn01,Schori02,Bowen03} which have dealt
with only frequency sidebands a few MHz apart from the carrier frequency,
and therefore the previous works are not compatible with photon counting.
Our scheme would have advantages even over the time-domain pulsed
scheme~\cite{Wenger05} with respect to spatial modes. The well-defined
frequency and spatial modes are important for efficient scaling of a
quantum circuit via interference of several beams.

Moreover our scheme is compatible with CV entanglement distillation with
the photon subtraction operation~\cite{Browne03}. Combining the
entanglement distillation and teleportation would result in the improvement
of teleportation fidelity~\cite{Olivares03}. Our entangled state could
also be applicable to loophole-free tests of Bell's
inequalities~\cite{Nha04,Garcia04}.

\vspace{2ex}

For an ideal two-mode squeezed vacuum state with infinite squeezing, the
EPR correlations can be described by the Wigner function of the two modes
$A$ and $B$:
\begin{equation}
W_{\rm EPR} (x_{\rm A}, p_{\rm A}; x_{\rm B}, p_{\rm B}) = c \delta(x_{\rm
A}-x_{\rm B}) \delta(p_{\rm A}+p_{\rm B}), \label{eq-1}
\end{equation}
where $x,p$ are the quadrature amplitudes corresponding to the conjugate
quadrature operators $\hat{x},\hat{p}$ which obey the commutation relation
$[\hat{x},\hat{p}]=i/2$~($\hbar=1/2$) in analogy with the position and
momentum operators. This Wigner function expresses that the quadrature
amplitudes of the two modes are perfectly correlated ($x_{\rm A}=x_{\rm
B}$) and anti-correlated ($p_{\rm A}=-p_{\rm B}$), respectively. In the
real experiment where only finite squeezing is available, the state is no
longer the ideal EPR state and shows finite variances in $\langle [\Delta
(\hat{x}_{\mathrm{A}}-\hat{x}_{\mathrm{B}} )]^2\rangle$ and $\langle
[\Delta (\hat{p}_{\mathrm{A}} +\hat{p}_{\mathrm{B}})]^2\rangle$.

The two-mode squeezed vacuum generated by subthreshold OPOs has EPR
correlations within the variance spectra $S_x (\Omega), S_p (\Omega)$ as
usually measured in the frequency domain experiments. Here, $\Omega$ is a
sideband frequency around the fundamental wavelength. When measuring the CW
beams in the time domain we need to specify a certain temporal mode within
which the entangled quantum state is defined. Because of the broadband
correlations of the CW beams, there is a large amount of freedom in
choosing a mode which will suit an intended application and will show
significant correlation below the vacuum level. For instance, in a
teleportation experiment the temporal mode could be chosen to match that of
the input state.

A simple choice of temporal mode is the square filter of duration $T$.
Defining the filtered quadrature operators
\begin{equation}
\hat{x}^f = \frac{1}{\sqrt{T}}\int^T_0 \hat{x}(t) dt, \label{eq-dif}
\end{equation}
and similarly for $\hat{p}^f$, these operators obey the same commutation
relation, $[\hat{x}^f,\hat{p}^f]=i/2$. The EPR variances within this
temporal mode will be given by~\cite{Sasaki06}
\begin{equation}
\langle [\Delta (\hat{x}^f_{\mathrm{A}}-\hat{x}^f_{\mathrm{B}} )]^2\rangle
= \frac{T}{2\pi}\int^{+\infty}_{-\infty} S_x (\Omega) \frac{\sin^2
(\frac{\Omega T}{2})}{(\frac{\Omega T}{2})^2} d\Omega,
\end{equation}
\begin{equation}
\langle [\Delta (\hat{p}^f_{\mathrm{A}}+\hat{p}^f_{\mathrm{B}} )]^2\rangle
= \frac{T}{2\pi}\int^{+\infty}_{-\infty} S_p (\Omega) \frac{\sin^2
(\frac{\Omega T}{2})}{(\frac{\Omega T}{2})^2} d\Omega.
\end{equation}
The variances are thus filtered by a sinc function, and by adjusting the
integration time $T$ the frequency range which contribute to the integrated
correlation can be selected. The square temporal filter is simple, but not
necessarily the most appropriate for an application of the time domain
entanglement. Theoretical investigations concerning the shape of the filter
combined with photon counting experiments have been done in
Refs.~\cite{Sasaki06,Molmer06}.

\vspace{2ex}

\begin{figure}[tbp]
\centerline{\scalebox{0.55}{\includegraphics{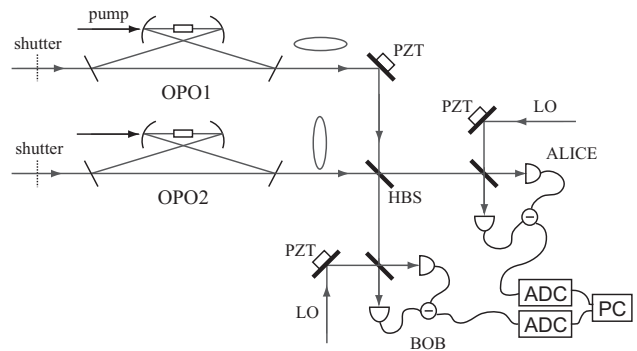}}} \caption{
Schematic setup of the experiment. OPOs: sub-threshold optical parametric
oscillators, HBS: a half beam-splitter, PZTs: piezo-electric transducers,
LOs: local oscillators, ADCs: an analogue-to-digital converter. }
\end{figure}

The schematic diagram of our experiment is illustrated in Fig. 1. This is
almost the same as the entanglement preparation part in our experiment of
quantum teleportation~\cite{Takei05}. The primary source of the experiment
is a CW Ti:Sapphire laser at 860nm, most of whose output is frequency
doubled in an external cavity. The output beam at 430nm is divided into two
beams to pump two OPOs.

A two-mode squeezed vacuum is produced by combining two squeezed vacua at a
half beam splitter~(HBS). Each squeezed vacuum is generated from a
subthreshold OPO with a 10-mm-long $\mathrm{KNbO_3}$ crystal. The crystal
is temperature-tuned for type-I noncritical phase matching. Each OPO cavity
is a bow-tie-type ring cavity consisting of two spherical mirrors (radius
of curvature 50mm) and two flat mirrors. The round trip length is about
500mm and the waist size in the crystal is 20$\mu$m. An output coupler has
transmissivity of 12.7\%, while the other mirrors are highly reflective
coated at 860nm. They have a high transmission for 430nm so that the pump
beam passes the crystal only once. The pump power is about 70mW for each
OPO. The total intracavity losses are around 2\%, giving a cavity bandwidth
of 7MHz HWHM. The resonant frequency of the OPO is locked via the FM
sideband locking method~\cite{Drever83}, by introducing a lock beam which
is propagating against the squeezed vacuum beam to avoid the interference
between the two beams. However a small fraction of the lock beam reflecting
from surfaces of the crystal circulates backward and contaminates the
squeezed state. This problem is resolved by changing the transverse mode
and the frequency of the lock beam, as described in detail in
Ref.~\cite{Polzik92a}.

The output beams from the HBS are sent to Alice and Bob, and then measured
using homodyne detectors with a bandwidth of 8.4 MHz. We lock relative
phases between EPR beams and local oscillators~(LOs) in the detection,
i.e., $x$ or $p$ quadrature, in the following way. First, weak coherent
beams which pass through the same paths as squeezed states are injected
into the OPOs from one of the flat mirrors and used for the conventional
dither and lock method. The relative phases between the weak coherent beams
and the squeezed vacua and between the weak coherent beams at the half beam
splitter (HBS) are actively controlled by applying feedback voltages to
piezos~(PZTs). The quadratures to be detected in the homodyne detectors are
also determined by locking the phases between the LOs and the injected
coherent beams. However these weak beams and the modulation on them
contaminate EPR correlation in some frequency ranges. Therefore we need to
remove such beams and lock the phases without the beams. To do so, we
introduce an electronic circuit that holds the feedback voltage and keeps
the phase relation for a while. Within 2msec, the beams are blocked with
mechanical shutters before the OPOs, and the EPR beams are measured with
the homodyne detectors. Each output from the detectors is sampled with an
analogue-to-digital converter~(ADC) at a rate of 50M samples per second and
then filtered on a PC to yield a number of measured quadrature values.

\begin{figure}[tbp]
\centerline{\scalebox{0.7}{\includegraphics{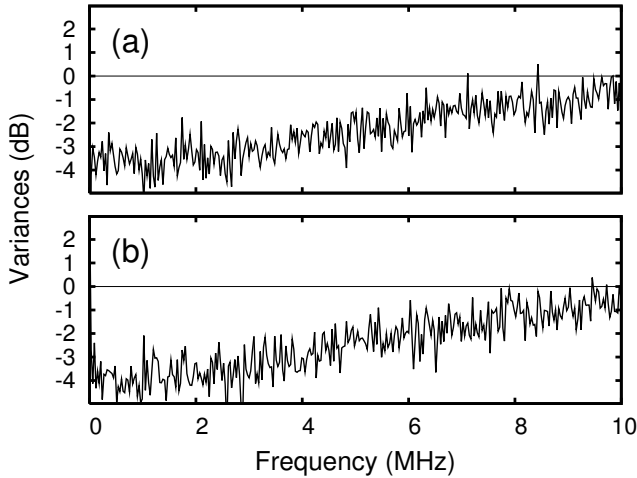}}} \caption{
Calculated Fourier analysis of the 50M-sampled raw data without average.
(a) $S_x (\Omega)$ for $\langle [\Delta
(\hat{x}^f_{\mathrm{A}}-\hat{x}^f_{\mathrm{B}} )]^2\rangle$. (b) $S_p (\Omega)$
for $\langle [\Delta (\hat{p}^f_{\mathrm{A}}
+\hat{p}^f_{\mathrm{B}})]^2\rangle$. Each trace is normalized to the
corresponding vacuum level. }
\end{figure}

\vspace{2ex}

Before observing the EPR correlation in the time domain, we estimate what
range of the frequency bandwidth can be used to yield a time-resolved
quadrature value from the OPOs. Figure 2 shows the frequency spectra $S_x
(\Omega)$ and $S_p (\Omega)$ for the measured EPR beams calculated by
performing Fourier transformation digitally on the 50M-sampled raw data.
The EPR correlation is observed over the full bandwidth of the OPOs. In
particular it is observed for low frequencies down to 5kHz (a high pass
filter with a cut-off of 5kHz is used to eliminate noise at frequencies
close to DC). The correlation at low frequencies is essential for the
photon subtraction experiment for non-Gaussian operation or entanglement
distillation \cite{comment}. From the results, we can define a quantum
state within a time interval of the inverse of the cavity bandwidth 7MHz.
But the EPR correlations degrade at higher frequencies, so we use the
temporal filter (2) with integration time $T=0.2\mu$sec, yielding 10000
points in each measurement. It follows from the frequency filters of sinc
fuctions in Eqs.~(3) and (4) that we select mainly a frequency range below 5MHz.

\begin{figure}[tbp]
\centerline{\scalebox{0.65}{\includegraphics{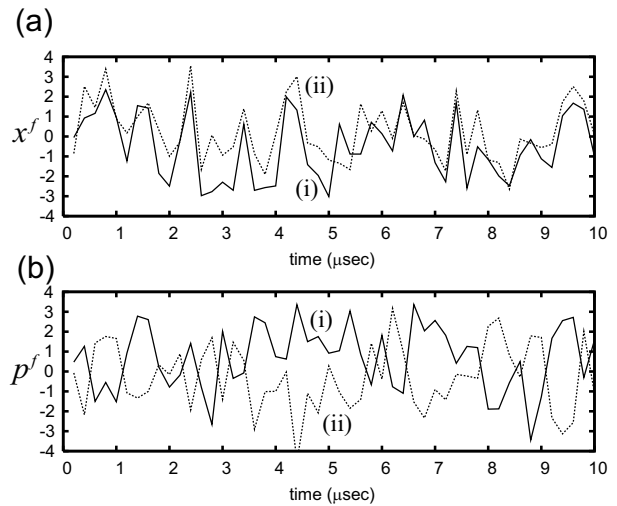}}} \caption{ Typical
measured correlations. Only 50 points are picked up. (a) and (b) are
measured quadrature values within time interval $T=0.2\mu$sec for $x$ and
$p$ quadratures, respectively. In each figure, trace (i) is for Alice,
while (ii) for Bob. }
\end{figure}

First we compare explicitly the time-resolved quadrature values measured at
Alice and Bob in every $0.2\mu$sec, rather than the variances of $\langle
[\Delta (\hat{x}^f_{{\rm A}}-\hat{x}^f_{{\rm B}} )]^2\rangle$ and $\langle
[\Delta (\hat{p}^f_{{\rm A}} +\hat{p}^f_{{\rm B}})]^2 \rangle$ performed in the
previous works~\cite{Ou92,Silberhorn01,Schori02,Bowen03}. This is also
different from the pulsed scheme~\cite{Wenger05} where EPR beams were
recombined at a beam splitter and returned to two squeezed vacua and then
one of which was measured in homodyne detection. Figure 3 shows the typical
example of the measured quadrature values within time interval
$T=0.2\mu$sec, where only 50 points are picked up. As mentioned above, the
measured values behave in such a way that $x$ and $p$ quadratures are
correlated ($x^f_{\rm A}\simeq x^f_{\rm B}$) and anticorrelated ($p^f_{\rm
A}\simeq -p^f_{\rm B}$), respectively. This manifests the originally devised
correlations discussed by EPR.

We then estimate the degree of EPR correlation using the correlation
diagrams with 10000 points as shown in Fig. 4. These results show clear
correlations in the relevant quadratures. By repeating the same measurement
ten times, we obtained the correlations of $\langle [\Delta (\hat{x}^f_{{\rm
A}}-\hat{x}^f_{{\rm B}} )]^2\rangle = -3.30\pm0.28{\rm dB}$ and $\langle
[\Delta (\hat{p}^f_{{\rm A}} +\hat{p}^f_{{\rm B}})]^2 \rangle=-3.74\pm0.32{\rm
dB}$ compared to the corresponding vacuum variances. Accordingly the
sufficient criteria~\cite{Duan00,Simon00} for quantum entanglement holds;
$\langle [\Delta (\hat{x}^f_{{\rm A}}-\hat{x}^f_{{\rm B}} )]^2\rangle +\langle
[\Delta (\hat{p}^f_{{\rm A}} +\hat{p}^f_{{\rm B}})]^2 \rangle=0.45\pm0.02<1$.
Therefore our generated state is entangled in the temporal mode defined
within $T=0.2\mu$sec. Note that we can decrease the integration time $T$
for fast data acquisition and QIP, while the EPR correlation will
 degrade due to the contribution of higher
frequency ranges. One would be able to obtain correlation over a broader
band by use of small-size OPOs with shorter round-trip length or waveguide
crystals, e.g., periodically poled Lithium Niobate waveguides
\cite{Yoshino}.

In summary we have experimentally demonstrated generation and
characterization of a two-mode squeezed vacuum state in the time domain. We
have observed the nonclassical correlation between the temporally measured
quadrature values. Our setup is almost the same as an entanglement
preparation part of our quantum teleportation setup~\cite{Takei05}. With
slight modification of classical channels, we will be able to perform the
broadband teleportation whereby it is possible to transfer quantum states
defined in the time domain like a Schr\"odinger cat-like state, a single
photon state and other non-Gaussian states. Moreover our scheme is
compatible with CV entanglement distillation with photon
subtraction~\cite{Browne03}. Our system has the well-defined frequency,
spatial and temporal modes. Therefore we could efficiently scale up quantum
circuits by interfering several beams toward the universal CV QIP.

\begin{figure}[tbp]
\centerline{\scalebox{0.5}{\includegraphics{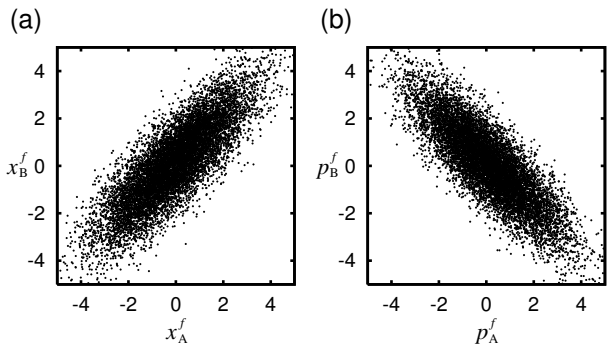}}}
\caption{Measured correlation diagrams of 10000 measured values for (a) $x$
and (b) $p$ quadratures.}
\end{figure}

This work was partly supported by the MEXT and the MPHPT of Japan, and
Research Foundation for Opto-Science and Technology.


\begin{thebibliography}{}
\bibitem{comment}Of course, it would be better if we could use very low frequencies near zero Hz. In practice, however, very low frequency components below 5kHz are often neglected in non-Gaussian state creation with photon subtraction \cite{Jonas06}. In that sense, we can also neglect very low frequency components for the application of teleporation of such non-Gaussian states.
\end{thebibliography}
\end{document}